\documentclass[]{article}
\newcommand{\bb}{\begin{eqnarray}}
\newcommand{\ee}{\end{eqnarray}}
\begin{document}

\centerline{\large\textbf{{BLACK HOLE ENTROPY}}} 
\centerline{P. Mitra}
\centerline{Saha Institute of Nuclear Physics}
\centerline{1/AF Bidhannagar, Calcutta 700064, India} 
\bigskip
%\date{arXiv:0902.2055}
\begin{abstract}
{We review black hole entropy with special reference to euclidean quantum
gravity, the brick wall approach and loop quantum gravity.}
\end{abstract}
%\bigskip\bigskip\bigskip\bigskip
\bigskip
\centerline{Talk given at  IAGRG meeting, Calcutta, January 28-31, 2009}
\bigskip%\newpage
\section{Introduction}
In Einstein's theory of gravitation, the gravitational field due
to a point mass is described by a metric which has many interesting
properties. Its black hole features have been known for a very
long time, but in the seventies it began to appear that thermodynamic
concepts like temperature and entropy were also associated with it.
Gradually it was realized that these were quantum effects. But the
degrees of freedom associated with the entropy could not be easily identified.
Many suggestions have been made: we shall discuss the entanglement
entropy approach and the more recent loop quantum gravity. 

After summarizing black hole mechanics, 
we consider the euclidean quantum gravity approach for both non-extremal
and extremal black holes.
There are indications of discontinuity between the two kinds, which 
arises in one way of quantization of the classical theory. An alternative way
leads to the Bekenstein-Hawking formula even for extremal black holes.

In order to understand the origin of black hole entropy, the entropy
of fields in black hole backgrounds was studied. This is identified
as entanglement entropy which arises because the region in the
interior of the horizon has to be traced over.

More recently, attempts have been made to formulate a quantum theory
of gravity itself. Black hole entropy has also been calculated in this
loop quantum gravity approach. This will be discussed in detail.

\section{Euclidean quantum gravity}
\subsection{Preliminaries}
A precursor of the idea of the entropy of black holes was the
{\bf area theorem}: the
area of the horizon of a system of black holes always increases
in a class of spacetimes. There were, 
more generally, a set of {\it laws of black hole mechanics} 
analogous to laws of thermodynamics:
\begin{itemize}
{\item zeroeth law: surface gravity $\kappa$
remains constant on the horizon of a black hole}
{\item first law:
${\kappa dA\over 8\pi}=dM-\phi dQ$,
where $A$ = area of horizon,
$\phi$ = potential at
horizon of black hole (with mass $M$, charge $Q$)}
{\item second law: the area of the horizon of a black hole system
always increases in spacetimes which are predictable from
partial Cauchy hypersurfaces..}
\end{itemize}
For charged black holes,
\bb
r_{\pm}=M\pm\sqrt{M^2-Q^2},
\ee
\bb
\kappa={r_+-r_-\over 2r_+^2},~ \phi={Q/r_+},~ A=4\pi r_+^2.
\ee

When   these   observations  were  made, there was no  obvious
connection with thermodynamics -- only a matter of  analogy.
But the existence of a
horizon  imposes    limitation  on the  amount  of  information
available  and hence may lead to an entropy, which should then be
measured by a geometric quantity associated with the  horizon,
namely its area. This implies,
$A\propto$ entropy, ${\kappa\over 8\pi}\propto$  temperature.

This interpretation was 
not fully convincing --
but quantum theory was found to cause dramatic changes in
black hole spacetimes: scalar field theory in a
Schwarzschild black hole background indicates the
radiation of particles at a temperature
\bb
T={\hbar\over 8\pi M}={\hbar\kappa\over 2\pi }.
\ee
This implies the connection  of the laws  of  black  hole
mechanics  with  thermodynamics, and  fixes a scale factor,
which involves Planck's constant indicating a quantum effect.

For Schwarzschild  black  holes, the first   law   of   {\it
thermodynamics} simplifies:
\bb
TdS=dM.
\ee
This can be integrated:
\bb
S={4\pi M^2\over \hbar}={A\over 4 \hbar}.
\ee

$T={\hbar\kappa\over 2\pi }$ is
generally valid for black holes having $g_{tt}\sim (1-
{r_h\over r})$. The first law becomes
\bb
T d{A\over 4\hbar}=dM-\phi dQ.
\ee
Comparison with the first law of {\it thermodynamics}
\bb
{T dS}=dM-\phi dQ
\ee
leads to the identification
\bb
S={A\over 4\hbar}.
\ee

\subsection{Non-extremal black holes}

A grand partition function may be written
for euclidean charged black holes:
\begin{eqnarray}
Z_{\rm grand}\equiv e^{-{M- TS-\phi Q\over T}}\approx e^{-I/\hbar}.
\end{eqnarray}
The functional  integral,  over  all   configurations  
(consistent with appropriate boundary conditions), 
is semiclassically approximated  by the maximum of the integrand.
The classical action $I$  can be calculated:
for a euclidean Reissner - Nordstr\"{o}m  black  hole  in a manifold
$\cal M$ with boundary which is subsequently taken to infinity,
\bb
I&=&-{1\over  16\pi}\int_{\cal M} d^4x\sqrt gR+{1\over 8\pi}\int_{\partial
{\cal M}} d^3x\sqrt\gamma (K-K_0)+\nonumber\\ &&{1\over 16\pi}\int_{\cal M}
d^4x\sqrt g F_{\mu\nu} F^{\mu\nu}.
\ee
Here $\gamma$ is the induced metric on the boundary $\partial {\cal M}$,
and $K$ the extrinsic curvature, from which a subtraction has to
be made.

The first  term  of the action  vanishes by Einstein's
equations ($R=0$).
To evaluate the second term, one takes the boundary of the manifold
at $r=r_B\to\infty$.
\bb
K&=&-{1\over\sqrt{g_{tt}}r^2}{1\over\sqrt{g_{rr}}}{d\over dr}
(\sqrt{g_{tt}}r^2)\nonumber\\ &=&-{1\over r^2}{d\over dr}
[(1-{M\over r}+\cdots)r^2]\nonumber\\ &=&-{1\over r^2}{d\over dr}
(r^2-Mr),
\ee
\bb
\int d^3x\sqrt\gamma=\int dt(1-{M\over r}+\cdots)4\pi r^2.
\ee
So $\int d^3x\sqrt\gamma K$ diverges as $r\to\infty$:
this can be cured by subtracting from $K$ its  flat  space
contribution  $K_0=-{1\over r^2}{d\over dr}r^2$. The
second piece of the action becomes
\bb
&&-{1\over 8\pi}\int dt(1-{M\over r}+\cdots)4\pi r^2 {1\over r^2}
{d\over dr}(-Mr)|_{r=r_B\to\infty} \nonumber\\ &=&-{1\over 2}\int dt (-M)
={1\over 2}\beta M.
\ee
The euclidean time $t$ has to go over one period
$0\to\beta %$\hbar/T=\hbar\beta$. 
=\frac{2\pi}{\kappa}$
to avoid a {\it conical singularity} at the horizon.

The third term  becomes
\bb
&&-{1\over 16\pi}\int dt .4\pi\int dr r^2. 2.{Q^2\over
r^4}\nonumber\\ &=&-{1\over 2}\int dt{Q^2\over r_+} \nonumber\\ &=&
-{1\over 2}\beta Q\phi,
\ee
where  $\phi$ is the electrostatic potential.
The negative sign is for the euclidean  solution.

Finally, 
\bb
I={1\over 2}\beta(M-Q\phi)={A\over 4},
\ee
\bb
M=T(S+{I\over \hbar})+\phi Q=T(S+{A\over 4\hbar})+\phi Q.
\ee
Now the Smarr formula reads
\bb
M={\kappa A\over 4\pi}+\phi Q
=T{A\over 2\hbar}+\phi Q,
\ee
implying $ S={A\over 4\hbar}$.

\subsection{Extremal black holes}
Extremal black holes have $r_+=r_-, Q=M, \phi=1$.
They are of special interest because the
topology changes discontinuously in the passage from the
(euclidean) non-extremal to the extremal case.

The action
\bb
I={1\over 2}\beta(M-Q\phi)=0,
\ee
\bb
M=T(S+{I\over \hbar})+\phi Q=TS+M\Rightarrow S=0
\ee
where $\beta$ has been assumed finite;
note that  $\lim_{Q\to M}\beta=\infty$
but there is no conical singularity in the extremal case, so 
there is no reason to fix the euclidean temperature, which can be
arbitrary.

Here, the quantum theory is based exclusively on
extremal topology. There is a more natural method of
quantization: {\it sum over topologies}.
Here the temperature $\beta^{-1}$ and the chemical
potential $\Phi$ are specified as inputs at the boundary $r_B$ of the manifold, 
while the mass $M$ and the charge $Q$ of the black hole are calculated
as functional integral averages. 
The definition of extremality $Q=M$ is imposed on these, making it a case of
{\it extremalization after
quantization}, as opposed to {\it quantization after
extremalization.}

A spherically symmetric class of metrics is considered;
the boundary conditions are:
\bb
g_{tt}(r_+)=0,~2\pi \sqrt{g_{tt}(r_B)}=\beta. 
\ee
\bb
A_t(r_+)=0, ~A_t(r_B)={\beta\Phi\over 2\pi i},
\ee
(the vector potential is taken to be zero).
Another boundary condition
reflects the extremal/non-extremal  nature:
\bb
\frac{1}{\sqrt{g_{rr}(r_+)}}\frac{d}{dr_+}\sqrt{g_{tt}(r_+)}
%{b'(r_+)\over\alpha(r_+)}
&=&1{\rm ...in~non-extremal~case},\nonumber\\
{\rm but~}&=&0 {\rm ...in~extremal~case}.
\ee

Variation of the action together with 
boundary  conditions  leads  to  reduced versions of
Einstein  -  Maxwell
equations,  whose solution
has a mass parameter $m$ and a charge $q$.
\bb
I&=& \beta(m-q\Phi) -\pi (m+\sqrt{m^2-q^2})^2
{\rm ~for~non-extremal~bc},\nonumber\\
I&=&\beta(m- q\Phi) {\rm ~for~extremal~bc}.
\ee
The partition function is of the form
\bb
\sum_{\rm topologies}\int d\mu(m)\int d\mu(q) e^{-I(q,m)},
\ee
with   $I$   appropriate for
non-extremal/extremal $q$.

The semiclassical approximation involves
replacing the double integral by the
maximum value of the integrand, {\it i.e.,} by 
$e^{-I_{min}}$,  where $I_{min}$ is the classical
action for the {\it non-extremal} case,
{\it minimized} with respect to $q,~m$,
yielding a function of $\beta,~\Phi$, and implying
$S=A/4$
for all values of $\beta, ~\Phi$.
The averages  $Q, ~M$, 
are calculated from $\beta,~\Phi$. The extremal limit
is reached for limiting values
\bb
\beta\to\infty,~~|\Phi|\to 1,~{\rm with}~\gamma\equiv\beta(1-|\Phi|)=
2\pi M ({\rm finite})
\ee
for the ensemble parameters $\beta,~\Phi$. Then
\bb
I={\gamma^2\over 4\pi}=\pi M^2,
\ee
\bb
Z\equiv e^{S-\gamma M/\hbar}=e^{-\pi M^2/\hbar},\ee
continuing to correspond to 
$S={A\over 4\hbar}$.

\section{Matter in black hole background}
To  study the entropy of a scalar field in the background 
provided by a black  hole, one may  employ   
brick-wall boundary conditions, 
where the wave function is cut off just outside the horizon
\begin{eqnarray}
\varphi(x)=0\qquad {\rm at}\;r=r_h+\epsilon
\end{eqnarray}
with $\epsilon$ an ultraviolet cut-off.
One also needs an infrared cut-off (box):
\begin{eqnarray}
\varphi(x)=0\qquad {\rm at}\;r=L>>r_h
\end{eqnarray}
We use a static, spherically symmetric black hole spacetime
\begin{eqnarray}
ds^2=g_{tt}(r)dt^2 +g_{rr}(r)dr^2+ g_{\theta\theta}(r)d\Omega^2.
\end{eqnarray}
An $r$- dependent  radial  wave  number is defined for
particles  with mass $m$, energy $E$ and orbital angular momentum
$l$:
\begin{eqnarray}
k_r^2(r,  l,  E)=  g_{rr}[-g^{tt} E^2 -
{l(l+1)g^{\theta\theta}} -m^2]\ge 0
\end{eqnarray}
One imposes on it the (semiclassical) quantization condition
\begin{eqnarray}
\frac{1}{\pi}\int_{r_h+\epsilon}^L~dr~k_r(r, l, E)=n_r~{\rm integral}.
\end{eqnarray}

The free energy $F$ at inverse temperature $\beta$
is given by a sum over single particle states:
\begin{eqnarray}
\beta F&=&\sum_{n_r, l, m_l}\log(1-e^{-\beta E})\nonumber \\
&\approx  &  \int  dl~(2l+1)\int  dn_r\log   (1-e^{-\beta   E})
\nonumber\\
&=&-\int  dl~(2l+1)\int d(\beta E)~(e^{\beta E} -1)^{-1} n_r \nonumber\\
&=& -{\beta\over\pi}\int  dl~(2l+1)
\int dE~(e^{\beta E} -1)^{-1}\int_{r_h+\epsilon}^L
dr~g_{rr}^{1/2}\nonumber\\
&& \sqrt{-g^{tt}E^2-{l(l+1)g^{\theta\theta}}-m^2} \nonumber\\
&=&  -{2\beta\over  3\pi}\int_{r_h+\epsilon}^L  dr~  g_{rr}^{1/
2}g_{\theta\theta}(-g_{tt})^{-3/2}
\nonumber\\&& \int dE~(e^{\beta E} -1)^{-1}
[E^2+g_{tt}m^2]^{3/2}.
\end{eqnarray}
The limits  of  integration  for  $l, E$ are such that
the arguments  of the square  roots  are   nonnegative.
$l$ integration is then explicit, while the
$E$ integral has to be approximated.
The contribution to the $r$ integral from  large $r$
is also present  in  flat spacetime:
\begin{eqnarray}
F_0=-{2\over 9\pi}L^3\int_m^\infty dE{(E^2-m^2)^{3/2} \over
e^{\beta E} -1} 
\end{eqnarray}
and is not relevant.
The contribution from small $r$ is  singular  in  the limit
$\epsilon\to  0$. 

For non-extremal black holes, $g_{rr}\propto (r-r_h)^{-1}$,
$g_{tt}\propto (r-r_h)$, while $g_{\theta\theta}$
is regular:
\begin{eqnarray}
F_{sing}\approx -{2\pi^3\over  45\epsilon\beta^4}
[(r-r_h)g_{rr}]^{1/2}(-{g_{tt}\over
r-r_h})^{-3/2}g_{\theta\theta}|_{r=r_h},
\end{eqnarray}
with corrections  involving  $m^2\beta^2$. The entropy
\begin{eqnarray}
S_{sing}= \beta^2 {\partial F_{sing}\over\partial\beta}= {8\pi^3\over  45\beta^3
\epsilon}
[(r-r_h)g_{rr}]^{1/2}(-{g_{tt}\over
r-r_h})^{-3/2}g_{\theta\theta}|_{r=r_h}.
\end{eqnarray}

Using the Hawking temperature
\begin{eqnarray}
{1\over\beta}&=&{1\over 2\pi}
(g_{rr})^{-1/ 2}{\partial \over\partial r}(-g_{tt})^{1/2}|_{r=r_h}
\nonumber\\ &=&{1\over 4\pi}(g_{rr})^{-1/ 2}(-g_{tt})^{-1/ 2}
{\partial \over\partial r}(-g_{tt})|_{r=r_h}
\nonumber\\ &= &{1\over 4\pi}[(r-r_h)g_{rr}]^{-1/ 2}(-{g_{tt}\over
r-r_h})^{1/2}|_{r=r_h}
\end{eqnarray}
and the proper radial width
(defined through $d\tilde r^2\equiv g_{rr}dr^2$)
\begin{eqnarray}
\tilde\epsilon=\tilde r(r_h+\epsilon)-\tilde r(r_h)
\approx 2\epsilon^{1/2}[(r-r_h)g_{rr}]^{1/2}|_{r=r_h},
\end{eqnarray}
\begin{eqnarray}
S_{sing}=  {1\over  90 \tilde\epsilon^2}
g_{\theta\theta}|_{r=r_h}=  {1\over  360\pi\tilde\epsilon^2}
{\rm Area}.
\end{eqnarray}

{This area factor crucially depends on the behaviour  of 
the metric near the horizon and is valid only for
non-extremal black holes; it
does {\bf not} emerge in  the extremal case.}

\section{Loop quantum gravity}

\subsection{Preliminaries}
This is an approach to a quantum theory of gravity called
loop quantum gravity or quantum geometry. 
A classical ``isolated horizon'' is the starting point: quantum states 
are built up by associating spin variables with ``punctures'' on 
this horizon. 
The entropy is obtained by counting all possible states 
consistent with a given area, more specifically, a particular
{\it eigenvalue of the area operator}.

A generic configuration may be taken to have $s_j$ punctures 
with spin $j,j=1/2,1,3/2....$ Then
\bb 2\sum_js_j\sqrt{j(j+1)}=A,\ee 
the horizon area in special units  
$$4\pi\gamma\ell_P^2=1,$$ 
where $\gamma$ is the so-called Barbero-Immirzi parameter and $\ell_P$ the Planck length.
There is a {\it spin projection constraint}
{$$\sum m=0,\quad {\rm over~all~punctures}$$}
$$m\in\{-j, -j+1,... j\} {\rm ~for~puncture~with~spin~}j.$$

\subsection{Spin 1/2}
For simplicity first consider spin 1/2 on {\it each} puncture.
The punctures have to be considered as distinguishable.
The number of punctures $n$ with spin 1/2 is given by
\bb 2n\sqrt{\frac34}=A,\ee
so {\it if we neglect the spin projection constraint}, the entropy
\bb n\ln 2={A\ln 2\over \sqrt 3}=
{A\ln 2\over 4\sqrt 3\pi\gamma\ell_P^2}.\ee
It involves $\gamma$, which can be chosen
to yield the Bekenstein-Hawking entropy:
\bb{A\over 4\ell_P^2}\Rightarrow
\gamma={\ln 2\over \sqrt 3\pi}.\ee

To implement the $m$ constraint,
the number $2^n$ of states is written as
$$2^n=1+{}^nC_2+...+{}^nC_n:$$
with the $(r+1)^{\rm th}$ term counting states with $r$ up spins. For zero total
projection, $m=0$, $n/2$ spins are up. If $n$ is odd, there is no such state,
but if $n$ is even, the number of
states =${}^nC_{n/2}$.
For large $n$, the Stirling approximation is 
\bb\ln n!&\simeq &\ln [\sqrt{2\pi n}({n\over e})^n]\nonumber\\
&=&n\ln n -n +\frac{1}{2}\ln (2\pi n),\nonumber\ee
$$\ln {n!\over (n/2)!(n/2)!}\simeq n\ln 2-\frac{1}{2}\ln n
+\frac{1}{2}\ln 2-\frac{1}{2}\ln\pi.$$
If the $n$ independent piece is neglected for large $n$, the entropy is 
\bb {A\over 4\ell_P^2}-\frac{1}{2}\ln A.
\ee
If in addition one wants the {\it total angular momentum to vanish},
the number of states with total projection 1 must be subtracted:
\bb{}^nC_{n/2}-{}^nC_{n/2+1}={}^nC_{n/2}(1-\frac{n/2}{n/2+1})
\Rightarrow{A\over 4\ell_P^2}-\frac{3}{2}\ln A.\ee

\subsection{General spin}
So far only $j=1/2$ spins have been considered at each puncture. 
If spin $j$ occur at all punctures,
an area $A$ needs $n=A/[2\sqrt{j(j+1)}]$ punctures.  
The number of states is $(2j+1)^n$ if the $m$ constraint is neglected.
This yields \bb n\ln (2j+1)=A\ln (2j+1)/[2\sqrt{j(j+1)}].\ee
This decreases with increasing $j$ (because $\ln (2j+1)$ increases
slowly compared with $\sqrt{j(j+1)}$).
Higher spins contribute less to entropy.
 
A general configuration may be taken to have $s_j$ punctures with spin $j$.
\bb N= {(\sum_j s_j)!\over \prod_j
s_j!}\prod_j (2j+1)^{s_j}\ee 
if the $m$ constraint is neglected
(the first factor gives the number of ways of choosing
locations of spins, the second factor counts
the numbers of spin states at the punctures).
One must sum $N$ over all nonnegative
$s_j$ consistent with a given $A$. We estimate the 
sum by maximizing $\ln N$ w.r.t. 
$s_j$ subject to fixed $A$.
Using Stirling again, and
neglecting the last piece therein, 
\bb\ln N=\sum_j s_j\ln
{2j+1\over s_j} + (\sum_j s_j)\ln (\sum_j s_j),\ee
\bb\delta\ln N=\sum_j\delta s_j\Big[\ln (2j+1)-\ln
s_j+\ln\sum_ks_k\Big],\ee
With some Lagrange multiplier
$\lambda$ to implement the area constraint, 
\bb\ln
(2j+1)-\ln s_j+\ln\sum_ks_k=\lambda\sqrt{j(j+1)}.\ee 
\bb
s_j=(2j+1)\exp\Big[-\lambda\sqrt{j(j+1)}\Big]\sum_ks_k\;.\ee 
Summing over $j$,
\bb\sum_j(2j+1)\exp\Big [-\lambda\sqrt{j(j+1)}\Big]=1,\ee 
which determines
$\lambda\approx$  1.72. 
\bb\ln N=\lambda A/2.\ee 
To make this ${A\over 4\ell_P^2}$ (with $4\pi\gamma \ell_P^2=1$)
the Barbero-Immirzi parameter becomes
\bb\gamma=\lambda /(2\pi)\approx 0.274.\ee 
Summing over $s_j$ may raise this 
value, while the projection constraint may lower it.

\subsection{Disregarding $j$ labels}

An alternative counting criterion regards 
states with the same $m$ on punctures
but having different $j$ to be equivalent, thus yielding fewer states.
It yields
\bb\sum_j (2+\delta_{j1})\exp [-\tilde\lambda\sqrt{j(j+1)}]=1.\ee
whence $\tilde\lambda\approx$ 1.58.
The rationale behind this criterion is supposed to be that $m$ is defined in a 
`surface Hilbert space' while $j$ 
is defined in a `volume Hilbert space'. However, the area of the 
horizon involves $j$ and both quantum numbers are associated with 
punctures on the horizon. As the area cannot be expressed in terms of
the $m$, this counting is more complicated and many 
attempts to implement it contain inaccuracies. This holds for
recipes as well as results given in the literature.

\subsection{Logarithmic corrections}

Let us impose the constraint of zero angular momentum projection.
Let $s_{j,m}$ punctures carry spin $j$ and projection $m$. Then
$s_j=\sum_ms_{j,m}$ is involved in the area constraint, while 
$\sum_{j,m}ms_{j,m}=0$.
The total number of ways of
distributing these spins 
\bb N_{\rm cor} 
={(\sum_js_j)!\over\prod_js_j!}\prod_j{s_j!\over\prod_ms_{j,m}!}=
{(\sum_{j,m}s_{j,m})!\over\prod_{j,m}s_{j,m}!}.\ee
To extremize the variation of $\ln N_{\rm cor}$ with two Lagrange multipliers
$\lambda,\alpha$ to implement the constraints, 
\bb
-\ln s_{j,m}+\ln\sum_{k,n}s_{k,n}=\lambda\sqrt{j(j+1)}+\alpha m,
\ee 
\bb
{s_{j,m}\over \sum_{k,n}s_{k,n}}=\exp[-\lambda\sqrt{j(j+1)}-\alpha m].
\ee
The projection constraint requires $\sum m\exp[-\alpha m]=0$,
i.e., $\alpha$=0, so that the distribution is the same as before. 

To estimate the sum over
$s_{j,m}$ configurations, one  
approximates the sum by an integral.
To study the variation of $\ln N_{\rm cor}$ with $s_{j,m}$, one notes that 
\bb\ln (s+\delta s)!&\simeq& (s+\delta s)\ln (s+\delta s)
- (s+\delta s)\nonumber\\
&\simeq& s\ln s -s +(\ln s)\delta s +(\delta s)^2/(2s).\nonumber\ee 
Terms linear in $\delta s$ cancel out because
of extremization, while the quadratic part
on exponentiation leads to factors of
$$\exp [-(\delta s_{j,m})^2/(2 s_{j,m})].$$ 
Factors of
$1/\sqrt{2\pi s_{j,m}}$ from Stirling's formula are cancelled
by $\sqrt{2\pi s_{j,m}}$ from gaussian integrations: 
\bb \int_{-\infty}^\infty d(\delta s_{j,m})\exp\Big
[-{(\delta s_{j,m})^2\over 2 s_{j,m}}\Big]=\sqrt{2\pi
s_{j,m}}.\ee 
Each  $\sqrt{s_{j,m}}\propto \sqrt{A}$.
The area constraint and the projection constraint
reduce the number of summations, hence reducing the number
of factors of  $\sqrt{A}$ by {\bf two}. 
But the numerator has an extra
factor $(\sum s_{j,m})^{1/2}\propto \sqrt{A}$. 
A net factor $1/\sqrt{A}$ survives, and the entropy 
\bb\ln  \sum N_{\rm cor}\simeq\lambda A/2-\frac12\ln A.\ee
This is the same log correction as for spin 1/2.

\subsection{Departure from linearity}
The linearity of $S$ with $A$ is not borne out by numerical 
investigations which fix the area sharply.
This can be understood by realizing the nature of the area constraint.
\bb A=2\sum_js_j\sqrt{j(j+1)}=[s_{1/2}\sqrt{3}+2s_{1}\sqrt{2}+s_{3/2}\sqrt{15}+\dots],\ee 
so if the natural numbers $s_j$ change, $s_{1/2}, s_{1}, s_{3/2},...$ cannot mix,
but some mixing is still possible:
\bb
A=&&[s_{1/2}\sqrt{3}+ 4s_{3}\sqrt{3}+15s_{25/2}\sqrt{3}+\dots]\nonumber\\
+&&[2s_{1}\sqrt{2}+12s_{8}\sqrt{2}+70s_{49}\sqrt{2}+\dots]+\cdots:
\ee
each set must be separately constant. One finds
sets of compatible spins $N_1\equiv\{1/2,3,25/2,...\}, N_2\equiv\{1,8,49,...\},...$.
There may be several constraints, {\it the number depending on the area,}
\bb A=\sum_NA_N\equiv 2\sum_N\sum_{j\in N}s_j\sqrt{j(j+1)}.\ee 
Corresponding Lagrange multipliers $\lambda_N$ satisfy 
\bb\sum_N\sum_{j\in N}(2j+1)\exp\Big [-\lambda_N\sqrt{j(j+1)}\Big]=1,\ee 
\bb S=\sum_N \lambda_NA_N/2=\bar\lambda A/2,\ee
$\bar\lambda\le 1.72$, depending on the ratios $A_1:A_2:\cdots$. $S$
reaches $1.72A/2$ only at special values of $A$ and is generally smaller.
%\end{document}

\newpage
%\vspace{-1.5 cm}

\end{document}